\documentclass{ws-procs975x65}		
\usepackage{grffile}
\newcommand \su[1] {\mathrm{SU}(#1)}

\newcommand \nubar {\overline{\nu}}

\newcommand \tra {\mathrm{Tr}}
\newcommand \mX[1] {m_{\textnormal{#1}}}

\begin{document}
\title{Recent results from $\su{2}$ with one adjoint Dirac fermion}

\author{A.~Athenodorou$^{ab}$, E.~Bennett$^{a*}$, G.~Bergner$^c$, and B.~Lucini$^a$}

\address{$^a$Department of Physics, Swansea University, Singleton Park, Swansea, SA2 8PP UK\\
$^*$E-mail: E.J.Bennett@swansea.ac.uk}

\address{$^b$Department of Physics, University of Cyprus, POB 20537, 1678 Nicosia, Cyprus}

\address{$^c$Universit\"at Bern, Institut f\"ur Theoretische Physik, Sidlerstr.\ 5, CH-3012 Bern, Switzerland}

\begin{abstract}
We present some results for $\su{2}$ with one adjoint Dirac flavour from lattice studies. Data for the spectroscopy, the static potential, topological charge, and the anomalous dimension of the fermionic condensate are included. Our findings are found to be inconsistent with conventional confining behaviour, instead pointing tentatively towards a theory lying within or very near the onset of the conformal window, with an anomalous dimension of the fermionic condensate of almost 1. Implications of these findings on the building of models of strongly-interacting dynamics beyond the standard model are discussed.
\end{abstract}


\bodymatter
\section{Introduction}
Novel strong interactions provide a promising explanation for the mechanism of electroweak symmetry breaking. Theories near the onset of the conformal window with an anomalous dimension of the chiral condensate of order one may in particular be relevant for phenomenology. It is important that these properties are verified for candidate models in a first-principles, fully non-perturbative fashion, and numerical lattice simulations provide an indispensable tool to this end. Recent lattice calculations \cite{Catterall:2007yx,DelDebbio:2009fd} have demonstrated that the $\su{2}$ gauge theory with two adjoint Dirac flavours is in the conformal window, while studies of many-flavour QCD \cite{aoki-scgt15,hasenfratz-scgt15,rinaldi-scgt15,schaich-scgt15} and of the $\su{3}$ gauge theory with two Dirac flavours in the sextet representation \cite{hasenfratz-poster-scgt15,kuti-scgt15,wong-scgt15} are ongoing.

In these proceedings we present results from an ongoing investigation of the $\su{2}$ Yang--Mills theory with one adjoint Dirac flavour, using Monte Carlo studies of the theory discretised on a spacetime lattice. The chiral symmetry breaking of this model produces only two Goldstone bosons, making it insufficient to give mass to the $W^\pm$ and $Z$ bosons. Thus the theory needs to be extended for phenomenological considerations. The main purposes of this study are firstly, to offer potential insights into the connection between near-conformal gauge theories and the presence of a large anomalous dimension. Secondly, to seek to pin down the lower end of the conformal window, which may give insight as to whether an interpolating theory between one and two adjoint Dirac flavours could be of phenomenological interest. Third, as an end rather than a means, as investigation and classification of gauge theories is also of theoretical interest.

\section{Lattice computations}
We simulate the theory on a 4-dimensional hypercubic lattice, using the Wilson gauge action and the Wilson fermionic action, giving\vspace{1pt}
\[
	S = \beta \sum_p \tra [1-U(p)] + \sum_{x,y} \overline{\psi}(x)D(x,y)\psi(y)\;,
\]
where the first sum is over plaquettes, and $D(x,y)$ is the massive Dirac operator:
\[
	D(x,y) = \delta_{x,y} - \kappa \left[\left(1-\gamma_\mu\right) U_\mu(x)\delta_{y,x+\mu} + \left(1+\gamma_\mu\right) U_\mu^\dagger(x-\mu)\delta_{y,x-\mu}\right]\;.
\]
$\kappa$ parameterises the bare fermion mass $am$, being related to it as $\kappa = 1/(8+2am)$. $\beta=2N/g^2=4/g^2$ is the inverse coupling, where $N=2$ is the dimension of the gauge group.

Gauge ensembles were generated using the HiRep code suite\cite{}, making use of the Rational Hybrid Monte Carlo (RHMC) algorithm. Further detail of the initial set of ensembles are given in \cite{Athenodorou:2014eua}; the newer ensembles introduced in this work were generated using the same procedures.

%
%
%
%

Ensembles were generated at two values of $\beta$: at $\beta=2.05$, masses of $-1.524 \le am \le -1.475$ were used, while at $\beta=2.2$ the range $-1.378 \le am \le -1.280$. With the lattice volume fixed to $N_T \times N_L^3$, spatial lattice sizes of 12, 16, and 24 were chosen, in all cases being cautious to avoid the effects of finite volume artefacts in our observables.

\begin{table}
\tbl{Parameter sets considered at $\beta= 2.05$ and $2.2$, showing their volumes, bare masses, PCAC masses, and number of configurations.}
{\null\hfill\begin{minipage}[t]{0.49\textwidth}
\begin{center}
(a) $\beta=2.05$
\begin{tabular}{@{}ccccc@{}}
\toprule
 & $N_L$ & $-am$ & $a\mX{PCAC}$ & $N_{\textnormal{conf}}$ \\\colrule
B1	&	12	 &	1.475 &	0.1493(5)&	1500  \\
B2	&	12	 &	1.500 &	0.1113(8)&	1500  \\
B3	&	12	 &	1.510 &	0.09226(92)&	4000  \\
C1	&	16	 &	1.475 &	0.1485(4)&	1500 \\
C2	&	16	 &	1.490 &	0.1279(2)&	1500   \\	
C3	&	16	 &	1.510 &	0.09111(31)&	1500 \\
C4	&	16	 &	1.510 &	0.09048(52)&	1500 \\
C5	&	16	 &	1.514 &	0.08223(34)&	1500 \\
C6	&	16	 &	1.519 &	0.06587(37)&	1500 \\
D1	&	24	 &	1.510 &	0.09130(27)&	1534 \\
D2	&	24	 &	1.523 &	0.04722(43)&	2168 \\
D3	&	24	 &	1.524 &	0.04081(71)&	800* \\\botrule
\end{tabular}
\end{center}
\end{minipage}
\hfill
\begin{minipage}[t]{0.49\textwidth}
\begin{center}
(b) $\beta=2.2$
\begin{tabular}{@{}ccccc@{}}
\toprule
 & $N_L$ & $-am$ & $a\mX{PCAC}$ & $N_{\textnormal{conf}}$ \\\colrule
2B1 &	12 &	1.280 &	0.2017(6) &	4000 \\
2B2 &	12 &	1.290 &	0.1891(5) &	4000 \\
2B3 &	12 &	1.300 &	0.1761(7) &	4000 \\
2B4 &	12 &	1.310 &	0.1633(7) &	4000 \\
2C1 &	16 &	1.320 &	0.1487(3) &	4000 \\
2C2 &	16 &	1.340 &	0.1189(4) &	4000 \\
2C3 &	16 &	1.350 &	0.1021(5) &	4000 \\
2C4 & 	16 &	1.360 &	0.0844(5) &	4000 \\
2D1 &	24 &	1.360 &	0.0830(10) &	1721 \\
2D2 &	24 & 1.370 &	0.0652(2) &	4000 \\
2D3 &	24 &	1.378 &	0.0486(5) &	1505 \\\botrule
\end{tabular}
\end{center}
\end{minipage}\hfill\null}

\begin{tabnote}
* Run D3 used configurations saved every 5 trajectories rather than every one, giving  equivalent statistics to 4000 trajectories.
\end{tabnote}
\label{tab:parameters}

\end{table}

Our aim in studying this theory is to probe its infrared regime---in particular, we would like to identify conformal properties and calculate the chiral condensate anomalous dimension. To this end, following \cite{DelDebbio:2010hu,DelDebbio:2010hx}, we have calculated the mass spectrum in the chiral regime, and tested finite-size scaling predictions. Additionally, we have performed fits of the Dirac mode number \cite{Patella:2012da}.

The massless fermionic action admits a global $\su{2}$ chiral symmetry, which is broken by a non-zero chiral condensate to $\mathrm{SO}(2)\equiv \mathrm{U}(1)$. The generator of the unbroken subgroup is identified with the baryon number $B$. Since parity is unbroken, the states of the theory are labeled by $B^P$, where $P$ is the parity, and by their spin. The spectrum comprises mesons, baryons, glueballs, and glue--fermion composite states \cite{Bergner:2013nwa}. In addition to these states, we also measure the string tension.

Mesonic states are accessed by the usual correlation functions of operators composed of fermion fields, which necessarily contain disconnected contributions due to the single flavour. Baryonic states meanwhile are accessed via the connected correlation functions of appropriately-chosen operators of the same form as for the mesons. Mesons have baryon number $B=0$, while baryons have $B=\pm2$. 

Meanwhile glueball operators consist of linear combinations of ordered products of link matrices on closed path, chosen to transform irreducibly under spin transformations, parity, and charge conjugation.

The string tension is accessed via two methods: via operators of blocked smeared Polyakov loops, fitted to an effective string theory prediction, and via the temporal asymptotic behaviour of expectation values of Wilson loops, giving the static potential, which can be fitted to a Cornell-like form. 

For further details of the methods used to generate the ensembles and to extract these observables, we defer to the more detailed presentation given in \cite{Athenodorou:2014eua}.

\def\figsubcap#1{\par\noindent\centering\footnotesize(#1)}
\begin{figure}
\begin{center}
\parbox{0.48\columnwidth}{\includegraphics[width=0.47\columnwidth]{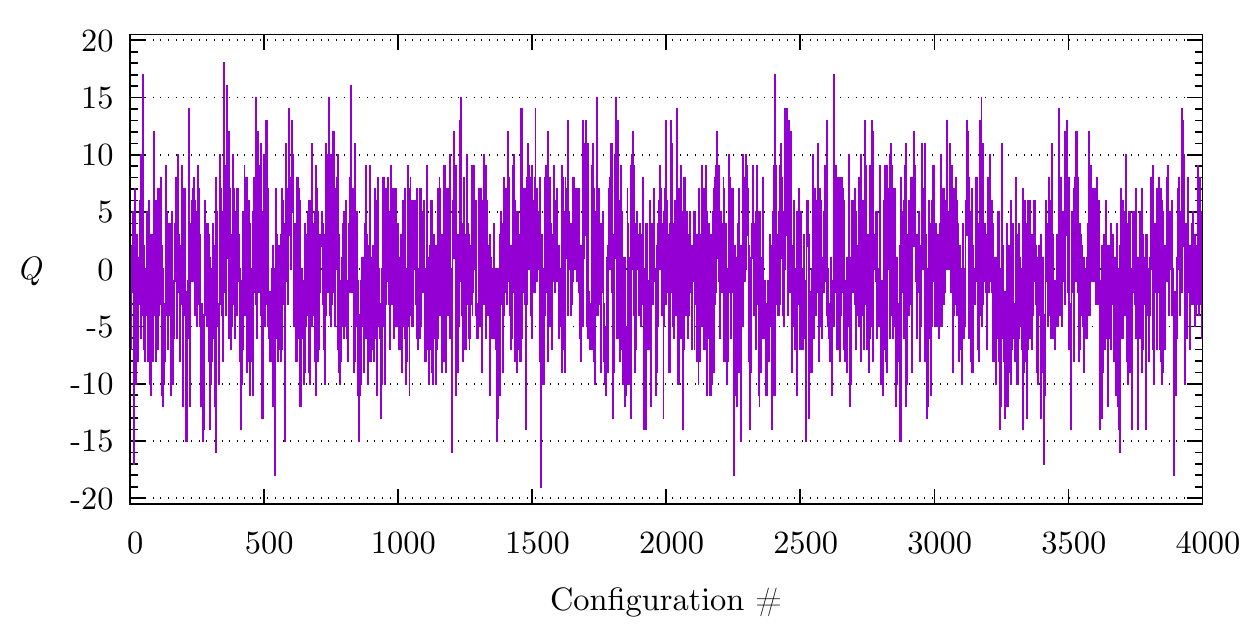}
\figsubcap{a}}
 \parbox{0.48\columnwidth}{\includegraphics[width=0.47\columnwidth]{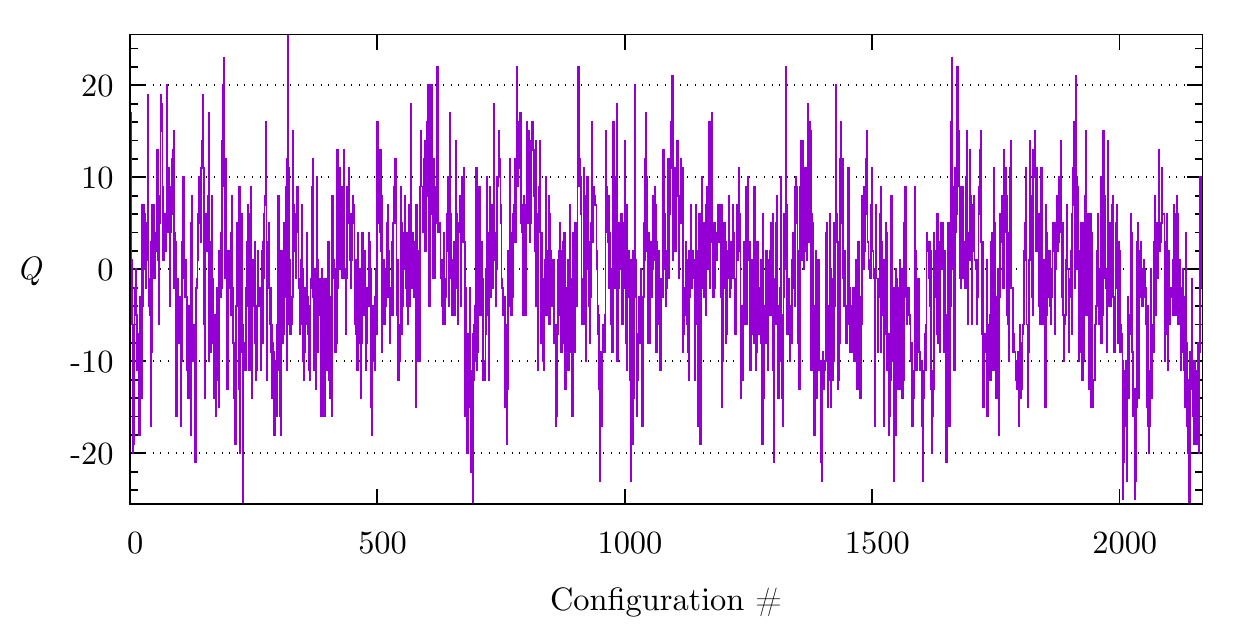}
 \figsubcap{b}}
 \caption{Sample topological charge histories for ensembles (a) C1, and (b) D2.}
\label{fig:Qhist}
\end{center}

\end{figure}

\begin{figure}
\begin{center}
\parbox{0.48\columnwidth}{\includegraphics[width=0.47\columnwidth]{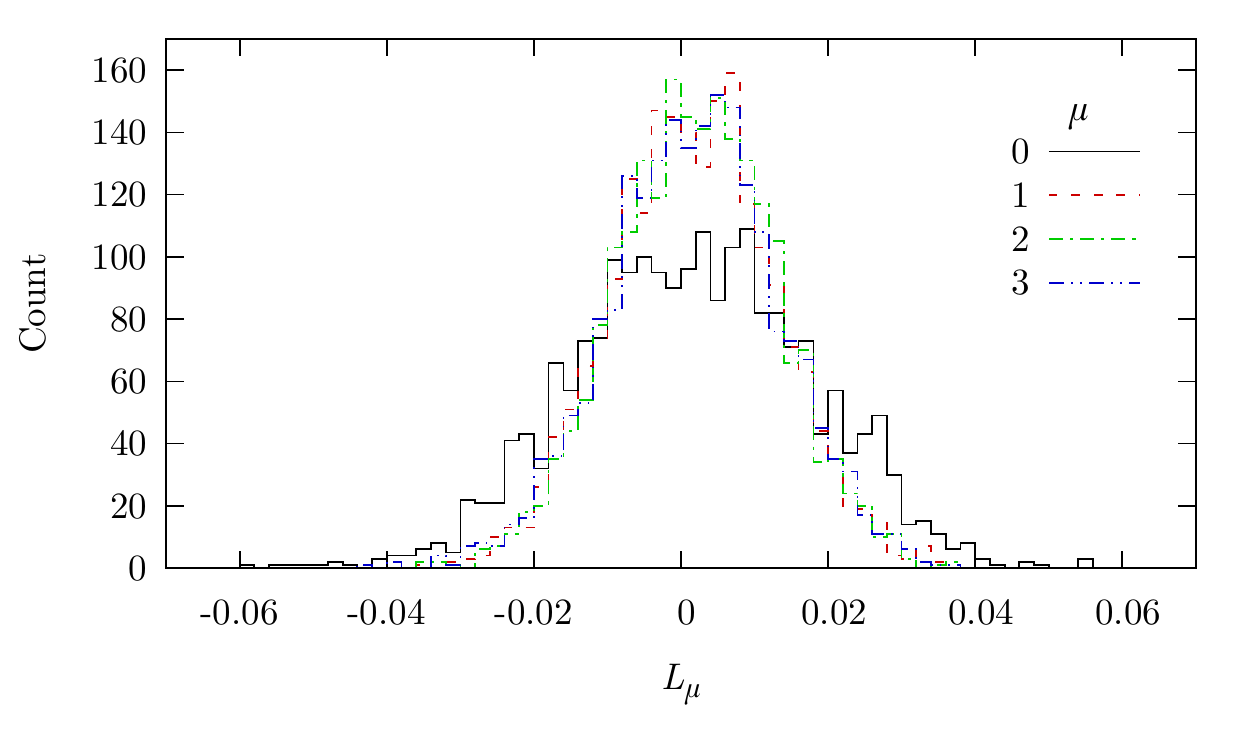}
\figsubcap{a}}
 \parbox{0.48\columnwidth}{\includegraphics[width=0.47\columnwidth]{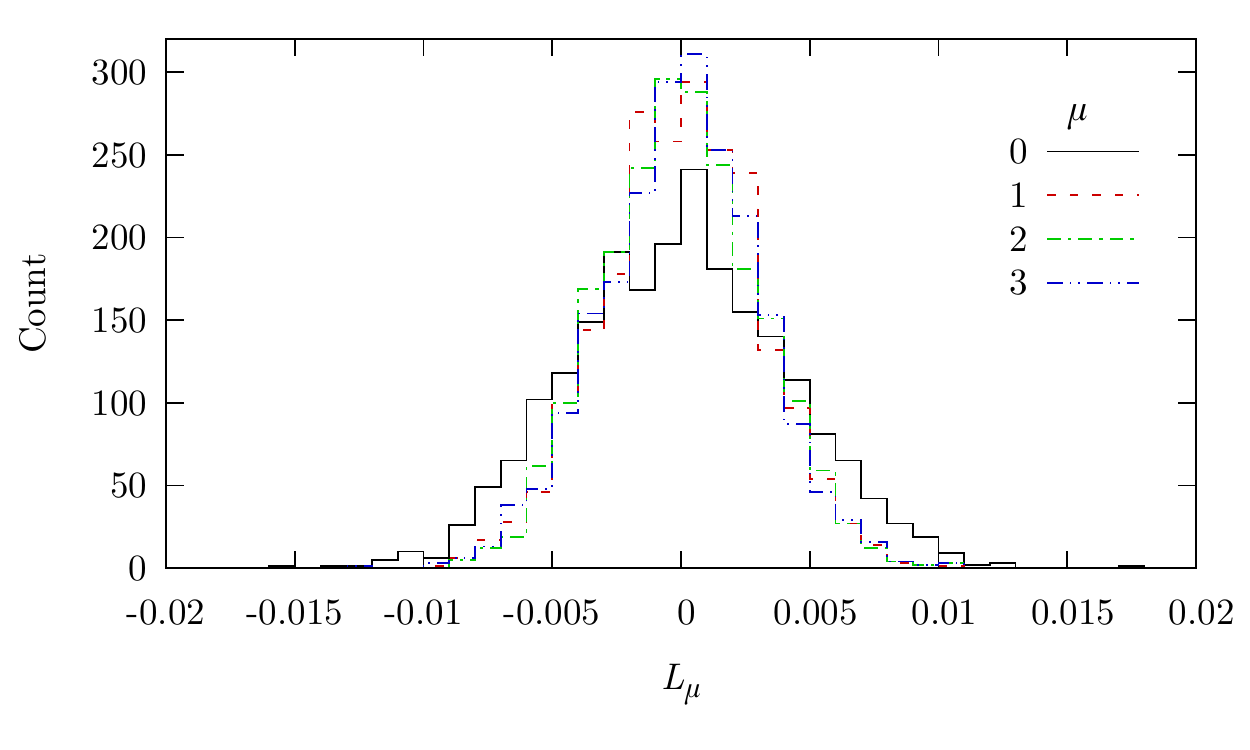}
 \figsubcap{b}}
 \caption{Polyakov loop distribution for ensembles (a) C4, and (b) D2.}
\label{fig:polyakov}
\end{center}

\end{figure}

As a check of ergodicity, we look at the topological charge history, illustrative examples of which are shown in Fig.~\ref{fig:Qhist}. Good ergodic movement is seen, with no sign of frozen regions. To verify that we are not in a region of lattice artefacts, the histogram of the average Polyakov loop in each direction is checked; in Fig.~\ref{fig:polyakov} we show illustrative examples showing an unbroken centre symmetry via the unimodal distribution---a broken centre would be indicated by a bimodal distribution.

\section{Spectrum}
\begin{figure}
\begin{center}
\parbox{0.48\columnwidth}{\includegraphics[width=0.47\columnwidth]{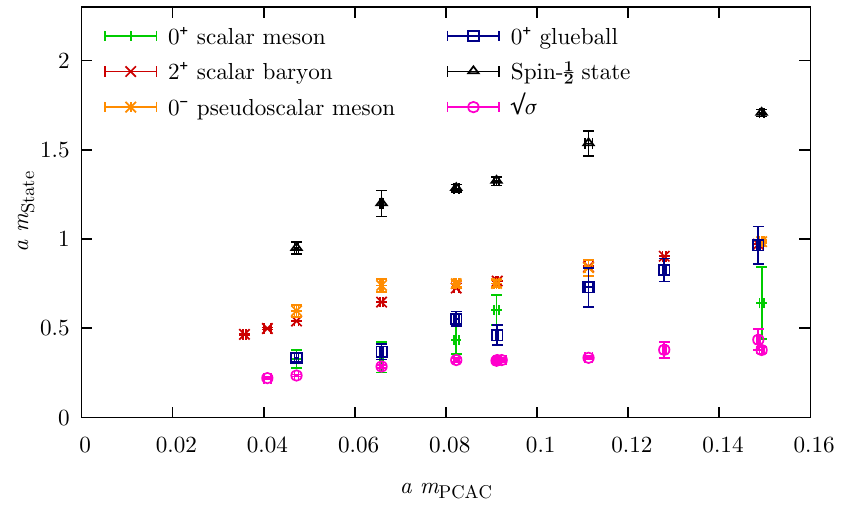}
\figsubcap{a}}
 \parbox{0.48\columnwidth}{\includegraphics[width=0.47\columnwidth]{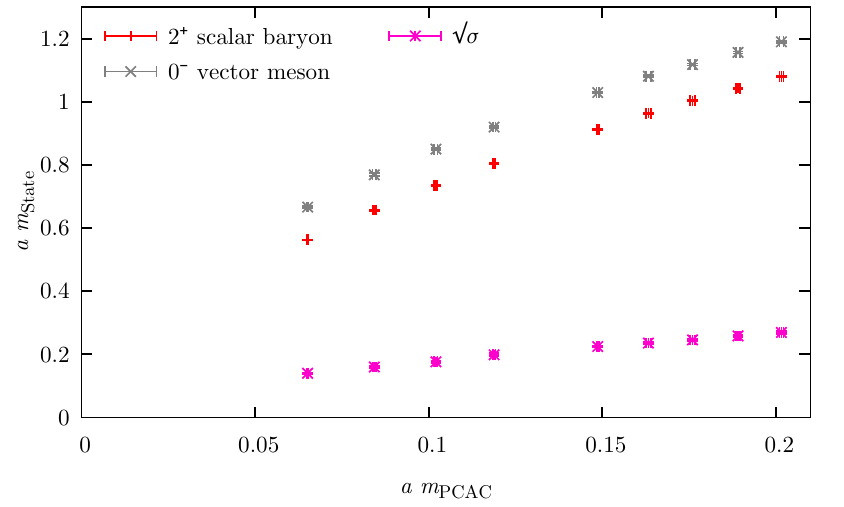}
 \figsubcap{b}}
 \caption{Bare spectral quantities for (a) $\beta=2.05$, and (b) $\beta=2.2$, as a function of  $a\mX{PCAC}$.}
\label{fig:spectrum-unscaled}
\end{center}

\end{figure}

\begin{figure}
\begin{center}
\parbox{0.48\columnwidth}{\includegraphics[width=0.47\columnwidth]{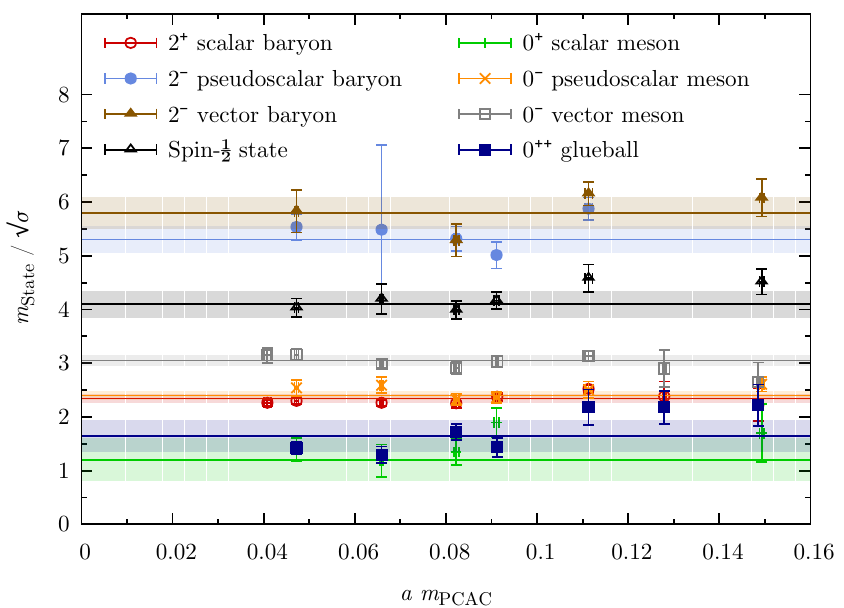}
\figsubcap{a}}
 \parbox{0.48\columnwidth}{\includegraphics[width=0.47\columnwidth]{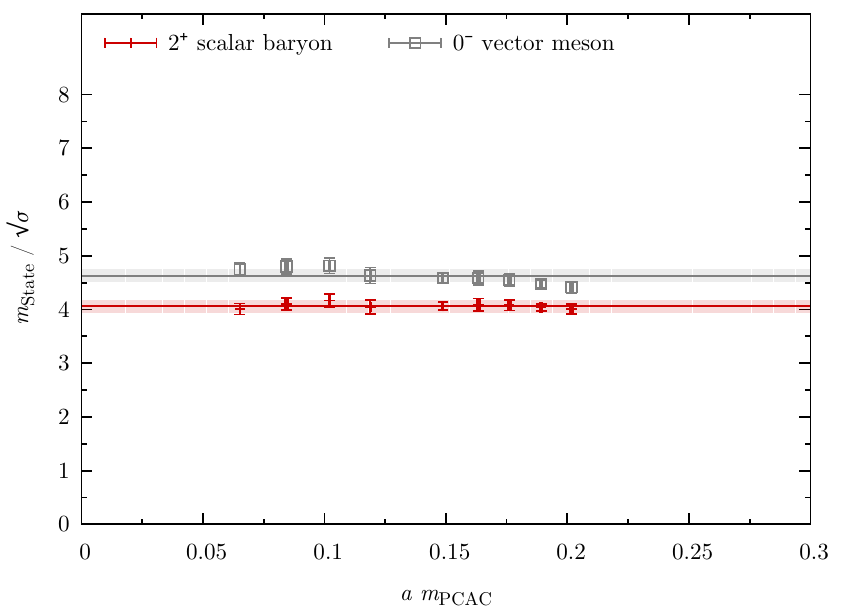}
 \figsubcap{b}}
 \caption{Spectral quantities for (a) $\beta=2.05$ and (b) $\beta=2.2$, scaled by the string tension $\sqrt{\sigma}$, as a function of the PCAC mass $a\mX{PCAC}$. In all cases the behaviour observed is consistent with flat.}
\label{fig:spectrum-scaled}
\end{center}

\end{figure}

We now turn our attention to the spectrum of the theory. As seen in Fig.~\ref{fig:spectrum-unscaled}, all spectral observables of the theory, including meson, baryon, and glueball masses, and the string tension in lattice units, decrease monotonically towards zero as $a\mX{PCAC}\rightarrow0$. 

In Fig.~\ref{fig:spectrum-scaled} we show the spectrum of the theory, now as dimensionless ratios to the string tension, both for $\beta=2.05$ and with available preliminary data for $\beta=2.2$. All states show constant mass ratios to two standard deviations throughout the range in which they were observed. At $\beta=2.05$, the $0^+$ scalar meson is seen to be the lightest state in the scaling region, and appears to be degenerate with the $0^{++}$ glueball, with which it is anticipated to mix.

\section{Chiral condensate anomalous dimension}
\begin{figure}
\begin{center}

\parbox{0.32\columnwidth}{\includegraphics[width=0.32\columnwidth]{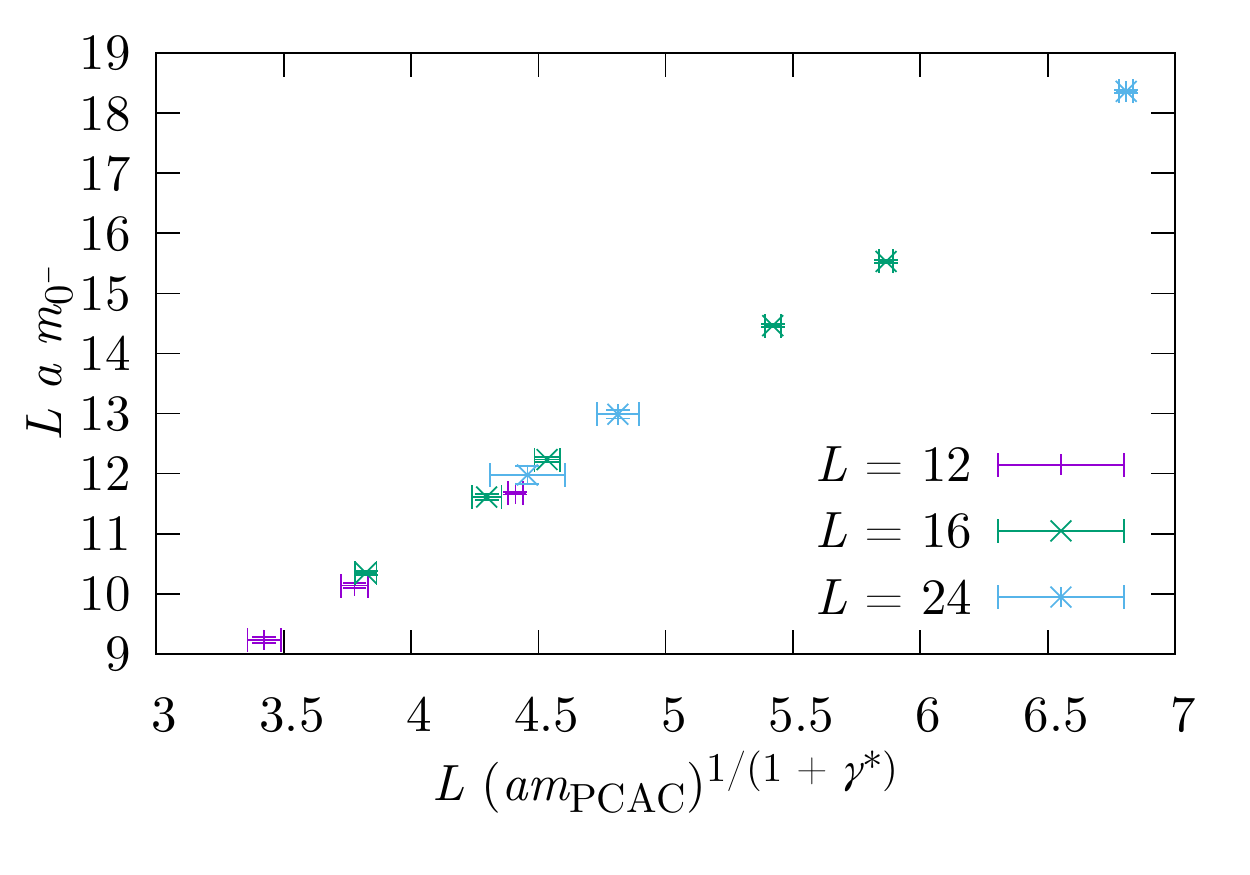}
\figsubcap{$\gamma_*=0.9$}}
 \parbox{0.33\columnwidth}{\includegraphics[width=0.33\columnwidth]{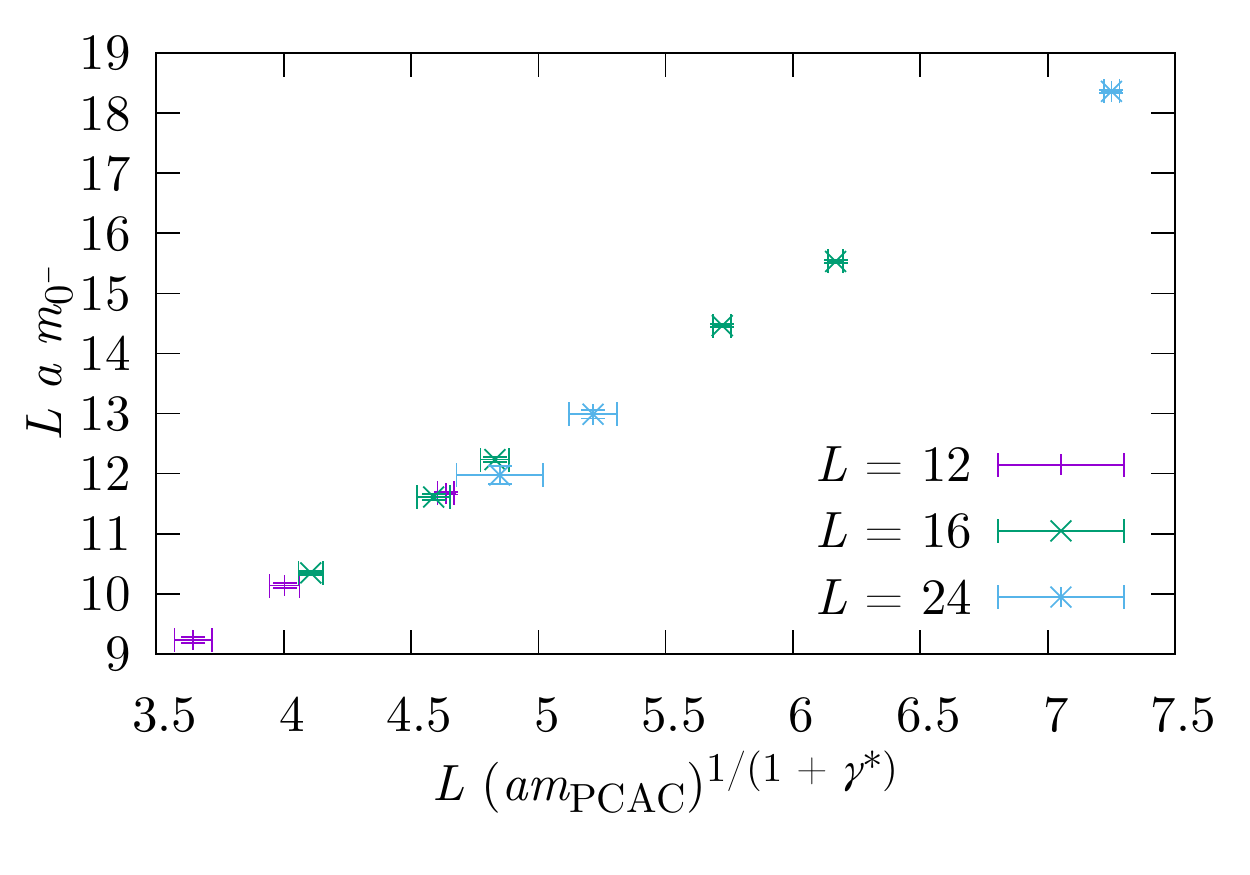}
 \figsubcap{$\gamma_*=1.0$}}
 \parbox{0.33\columnwidth}{\includegraphics[width=0.33\columnwidth]{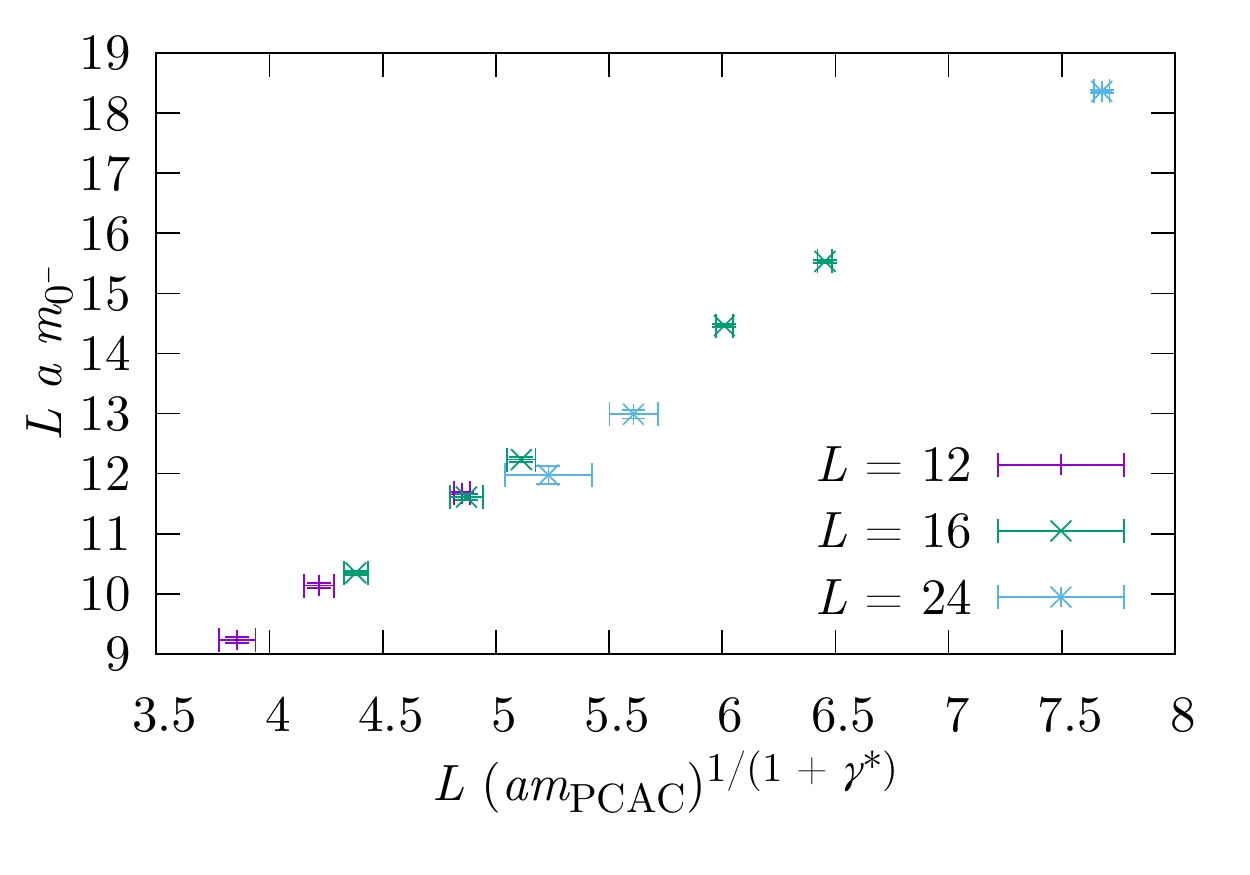}
 \figsubcap{$\gamma_*=1.1$}}
 \caption{Plots of $Lm_{0^-}$ as a function of $L\mX{PCAC}^{1/(1+\gamma_*)}$ for $\beta=2.05$ for three values of $\gamma_*$. The results appear to identify a universal curve for $0.9\le\gamma_*\le1.0$.}
\label{fig:scaling}
\end{center}

\end{figure}

We consider two independent methods to extract the anomalous dimension $\gamma_*$. The first gives an approximate value via finite size scaling predictions. For a conformal theory, a spectral quantity $\mX{X}$ of the system in a finite volume of spatial extent $L$, as $L\rightarrow\infty$ and the combination $L\mX{PCAC}^{1/(1+\gamma_*)}$ is kept constant, obeys the relation
\[
	L\mX{X} = f\left( L\mX{PCAC}^{\frac{1}{1+\gamma_*}}\right)\;,
\]
for some unknown function $f$. Thus by plotting $L\mX{X}$ as a function of $L\mX{PCAC}^{1/(1+\gamma_*)}$ for our data at various volumes, we may extract $\gamma_*$ by tuning it until the data lie on a single universal curve. 

A sample of such plots are shown in figure \ref{fig:scaling}; from this we can see a best fit in the range $0.9\le \gamma_* \le 1.0$.

\begin{figure}
\begin{center}

\hfill\includegraphics[width=0.46\textwidth]{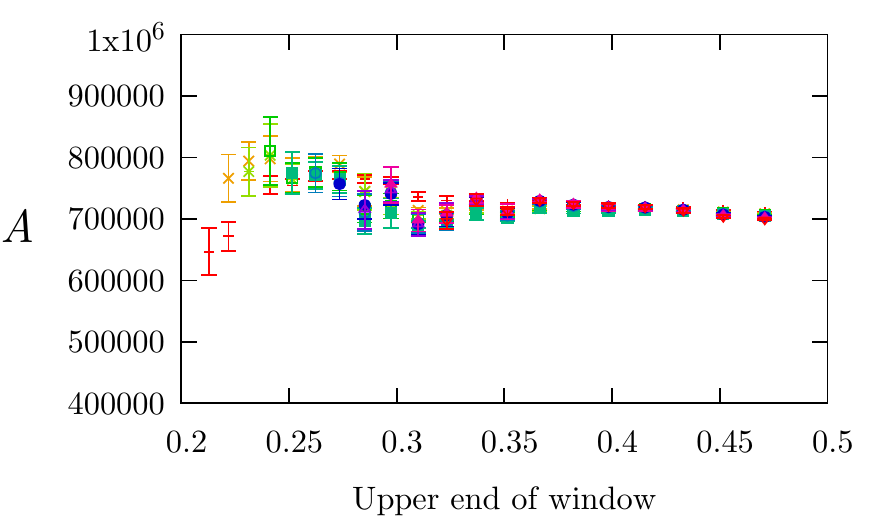}\hfill\includegraphics[width=0.46\textwidth]{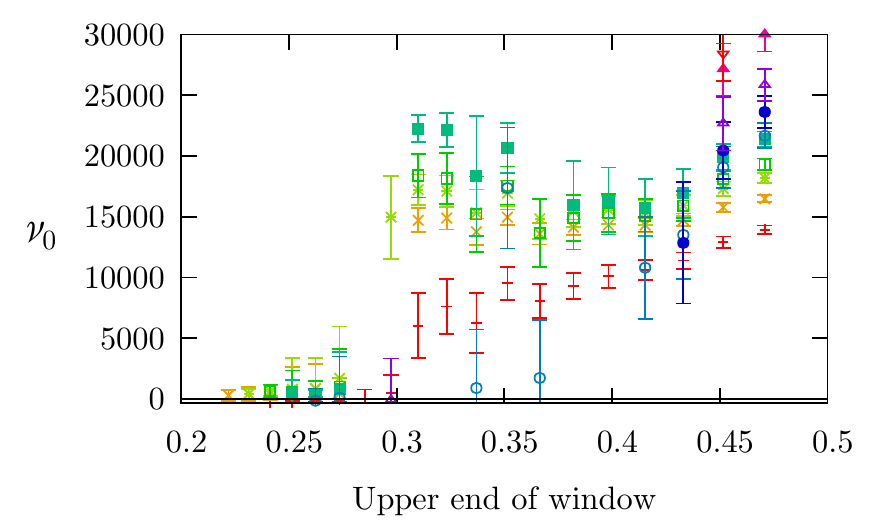}\hfill\null

\hfill\includegraphics[width=0.46\textwidth]{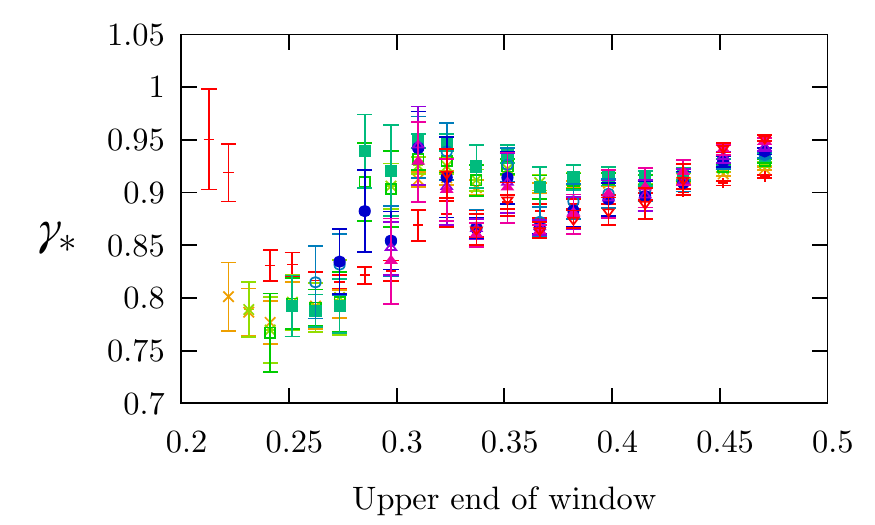}\hfill\includegraphics[width=0.46\textwidth]{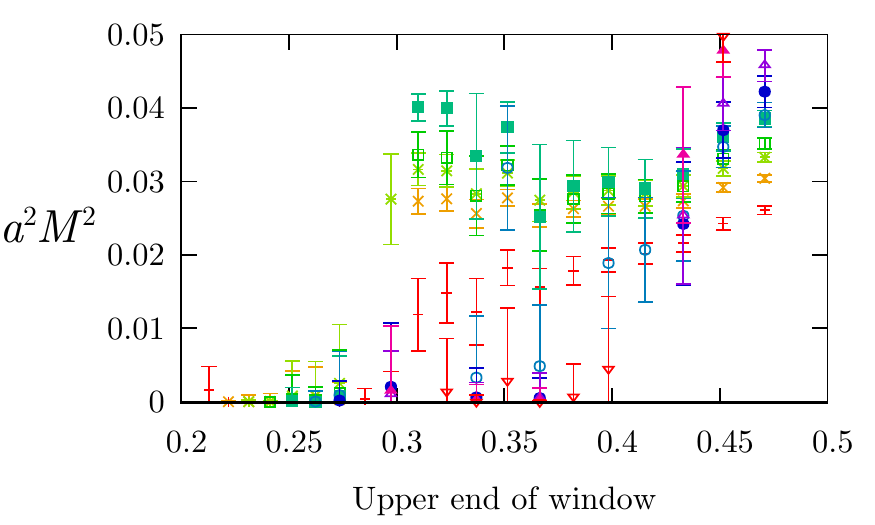}\hfill\null

\scriptsize{Lower end of window:}
\vspace{4pt}

\includegraphics[width=\textwidth]{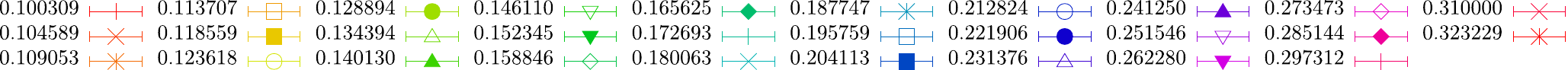}
 \caption{Plots of the fitted parameters $A$, $\nubar_0$, $\gamma_*$ and $m^2$ as a function of the window bounds. A region of stability is identified around $\gamma_*=0.92(1)$.}
 
\label{fig:modenumber}
\end{center}
\end{figure}

The second route we use to obtaining $\gamma_*$ is by fitting the Dirac mode number $\nubar(\Omega)$ as a function of the Dirac eigenvalues $\Omega$ \cite{Patella:2012da}. Since at low $\Omega$ the distribution is distorted by the finite fermion mass, while at large $\Omega$ the scaling region ends, we look for an ensemble with a large intermediary region to examine; thus a large lattice and small mass makes sense. We therefore examine the D2 ensemble, and look to fit
\[
	a^{-4}\nubar(\Omega) \approx a^{-4} \nubar_0(m) + A\left[(a\Omega)^2 - (aM)^2\right]^{\frac{2}{1+\gamma_*}}\;.
\]
Since $M$ here is a renormalised mass and not known \emph{a priori}, we have four parameters to fit: $\nubar_0$, $A$, $(aM)^2$, and $\gamma_*$, However, since the scaling region is also not known \emph{a priori}, an additional fitting degree of freedom is introduced in the choice of fitting window. Since the fit is highly sensitive to initial conditions, each possible window is fitted many times for varying starting parameters, and uncertainties estimated using the bootstrap method. We then look for a region where the fit is stable for a range of window lengths and positions, as can be seen in Fig.~\ref{fig:modenumber}; we see this gives a best fit of the anomalous dimension of $0.92(1)$. Further details of the fitting procedure are given in \cite{Athenodorou:2014eua}.

\section{Conclusions}
We have extended our previous investigation of $\su{2}$ with one adjoint Dirac flavour to a second value of $\beta$, where we have a preliminary set of observables available. At both values of $\beta$ we see behaviour consistent with conformality and inconsistent with QCD-like confinement: namely, monotonic change of all spectral quantities, whose ratios are flat, consistent with hyperscaling behaviour. At one value of $\beta$, we find a large anomalous dimension of the chiral condensate, $0.9\le \gamma_*\le0.95$, with a best fit of $0.92(1)$. Work is ongoing to extend the study at $\beta=2.2$ to the full set of observables studied for $\beta=2.05$---including the anomalous dimension of the chiral condensate, which we currently have insufficient data to determine conclusively---after which we intend to study sufficiently many values of $\beta$ to allow a continuum limit extrapolation.

\section*{Acknowledgements}
Computational resources used in this work include the Dirac and Dirac2 facilities, and the Blue Joule machine at the Hartree Centre, supported by STFC, and the HPC Wales cluster supported through WEFO.

\bibliographystyle{ws-procs975x65}
\bibliography{ws-procs975x65}

\end{document}